\pdfoutput=1
\documentclass[12pt]{article}
\usepackage[utf8]{inputenc}
\usepackage[bookmarks=true]{hyperref}
\usepackage{graphicx,amssymb,amsmath}
\usepackage{color,graphicx}
\usepackage{cite}
\usepackage[top=2.5cm, bottom=2.5cm, left=2.5cm, right=2.5cm]{geometry}	

\newcommand{\mt}[1]{\textrm{\scriptsize #1}}
\def\Nf{N_\mt{f}}
\def\Nc{N_\mt{c}}

\begin{document}

\begin{center}

\centering{\Large {\bf NICER view on holographic QCD}}

\vspace{5mm}

\renewcommand\thefootnote{\mbox{$\fnsymbol{footnote}$}}
Niko Jokela$\footnote{niko.jokela@helsinki.fi}$

\vspace{2mm}
{\small \sl Department of Physics} and {\small \sl Helsinki Institute of Physics} \\
{\small \sl P.O.Box 64} \\
{\small \sl FIN-00014 University of Helsinki, Finland}

\end{center}


\setcounter{footnote}{0}
\renewcommand\thefootnote{\mbox{\arabic{footnote}}}

\begin{abstract}
\noindent
The holographic models for dense QCD matter work surprisingly well. A general implication seems that the deconfinement phase transition dictates the maximum mass of neutron stars. The nuclear matter phase turns out to be rather stiff which, if continuously merged with nuclear matter models based on effective field theories, leads to the conclusion that neutron stars do not have quark matter cores in the light of all current astrophysical data. We comment that as the perturbative QCD results are in stark contrast with strong coupling results, any future simulations of neutron star mergers incorporating corrections beyond ideal fluid should proceed cautiously. For this purpose, we provide a model which treats nuclear and quark matter phases in a unified framework at strong coupling.\end{abstract}


\section{Introduction}\label{intro}

The QCD phase diagram is theoretically in most parts unknown. This is because the equations are too hard to solve using traditional perturbative methods and the lattice formulation is still insufficient to map out dense regimes. At the advent of multimessenger era we are unraveling, the astrophysical input has already given us hints on how the matter behaves under extreme conditions those where the matter is in the verge of collapsing into black holes. We direly need microscopic understanding of this behavior, which might be encoded in the equation of state (EoS). 

We attempt to solve a QCD-like theory at strong coupling and in the Veneziano limit of large number of colors $\Nc\to\infty$ and flavors $\Nf\to\infty$, with $\Nc/\Nf=1$ using holographic duality \cite{Ramallo:2013bua}. We will review recent progress both for the equilibrium quantities aka equation of state as well as transport quantities near equilibrium. We show that the holographic model, anchored with known QCD physics at low densities has predictive power when extrapolated at finite densities and low temperatures. The conclusions we draw are in stark contrast with perturbative QCD and, in particular, we underline the observation that even with all astrophysical constraints taken into account, the neutron stars do not have quark matter cores. Interestingly, this is at tension with recent claims on the contrary \cite{Annala:2019puf} and begs for further study.

This talk is mainly based on articles \cite{Jokela:2018ers,Jokela:2020piw} as well as on work in progress \cite{toappear} dealing with holographic applications in neutron stars in sections~\ref{sec:equilibrium} and \ref{sec:applications} and on articles \cite{Hoyos:2020hmq,Hoyos:2021njg} describing transport properties of strongly interacting quark matter phase discussed in section~\ref{sec:inequilibrium}.

\section{Equilibrium}\label{sec:equilibrium}

We begin with a brief discussion of the holographic framework which we use to model QCD-like matter. This framework is based on a holographic model called V-QCD \cite{Jarvinen:2011qe}. Here the letter V stands for the Veneziano limit, $\Nc\to\infty,\Nf\to\infty$ while keeping their ratio fixed. In order to make contact with neutron stars we will in the end extrapolate down to $\Nc\to 3,\Nf\to 3$, since charm quarks are too heavy to be expected in neutron star densities. This model has since then been extended in many directions, notably to finite density and temperature \cite{Alho:2012mh,Alho:2013hsa}. 

This is a bottom-up approach in holography, but our model is loosely based on string theory construction, so we aim to follow the rules of string theory as closely as possible. The idea is that we have introduced many new parameters, essentially parametrizing the ignorance by potentials for various bulk fields, the metric, the dilaton, and the (temporal part of) the gauge potential in the search for finite density solutions. The potentials are not completely arbitrary as they have been chosen by various criteria, {\emph{e.g.}}, by demanding that the coupling constant (dual to the dilaton field) in the UV part of the theory is faithfully captured at two loops. Such constraints are robust enough that there is very little freedom in the parameter space left. The remaining freedom in the potentials are fixed by matching the thermodynamic quantities both for the gluon sector \cite{Gursoy:2009jd} and in the flavor sector to the 2+1 lattice QCD data \cite{Jokela:2018ers}. In addition to, {\emph{e.g.}}, reproducing the non-trivial bump in the interaction measure which comes out with no fine-tuning, in view towards finite density physics, the ability of V-QCD to match onto lattice data of the baryon number susceptibility is crucial. 

The comparison with QCD data is essential and the V-QCD works surprisingly well at low densities in the QGP phase.  Motivated by this success the model is then extrapolated to finite densities and very low temperatures, the environments in the interior of quiescent neutron stars. In addition to mimicking quark matter phase, there is also a baryonic nuclear matter phase. This phase emerges where expected, though, for simplicity we treat baryons in the homogeneous approximation \cite{Bergman:2007wp}. At low temperatures, which can essentially be thought to mean $T\approx 0$, the model then describes (unpaired) quark matter (deconfined and chirally symmetric as we assume massless quarks) phase at high density with a first order transition to holographic nuclear matter phase (chirally broken confined) and then a further transition to a confined chirally broken zero density phase \cite{Ishii:2019gta}. One remarkable feature of this framework is that it allows us to consider the same (holographic) model both for the nuclear and the quark matter phase; this is a clear advantage when we later discuss applications to neutron stars.

\begin{figure}[!ht]
\centering
\includegraphics[width=0.7\textwidth]{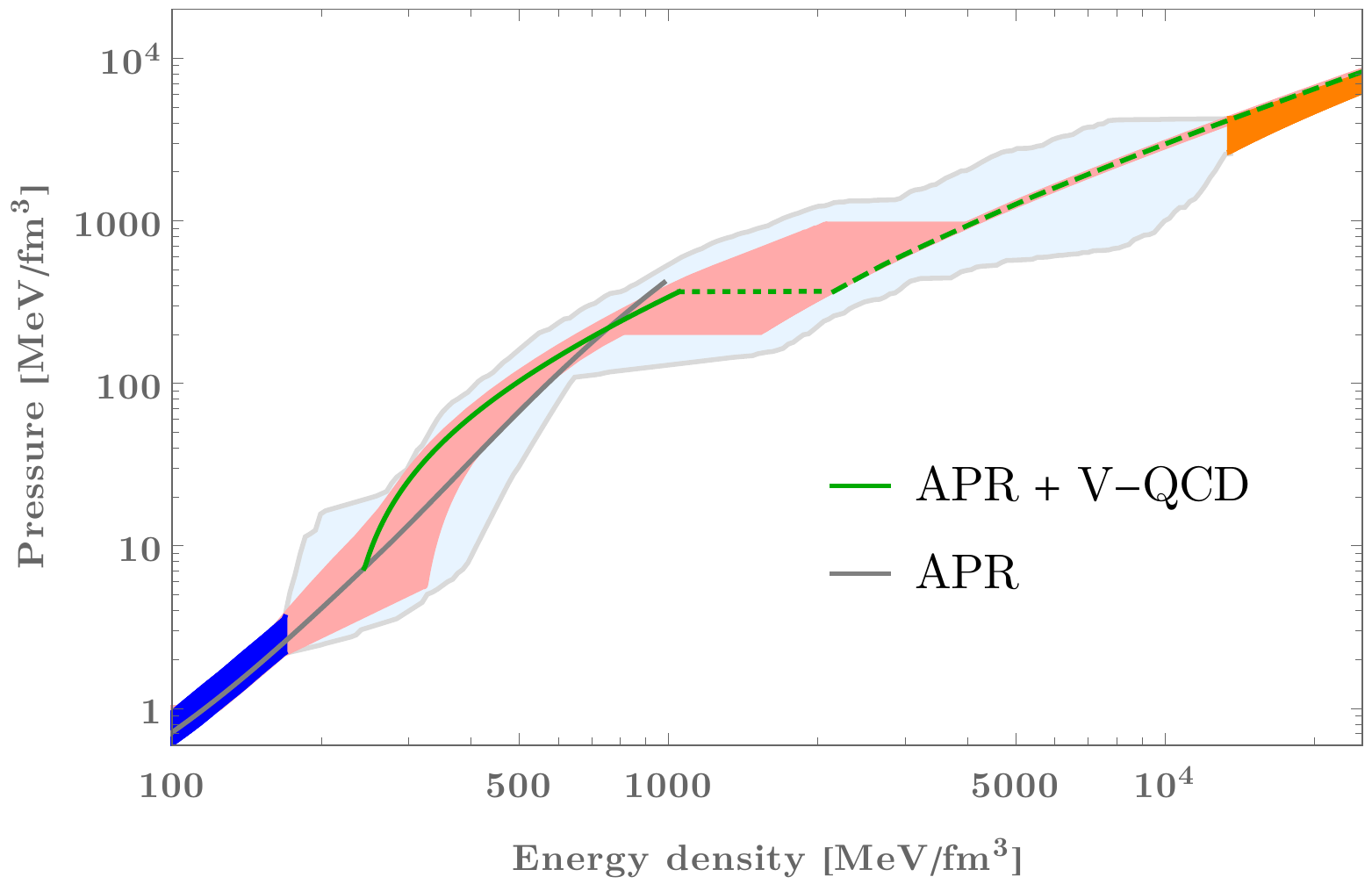}
 \vspace{-7.5cm} 
  \phantom{a}\hspace{4mm}\includegraphics[angle=0,width=0.7\textwidth]{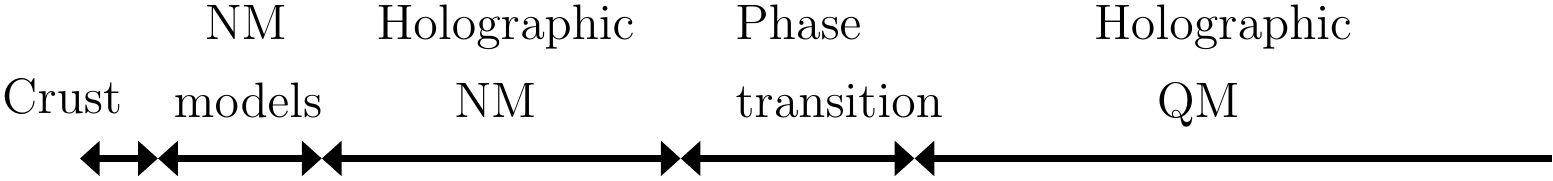}
 \vspace{7.cm}
\caption{The hybrid equations of state and constructed such that at low densities we adopt the nuclear matter models from literature and merge them into holographic V-QCD at densities $n\sim 1.2-2 n_s$ after of which it is just holography. It is noteworthy that the EoS from holography correctly matches onto pQCD constraints on the right, plotted in orange. As an example, V-QCD matched with APR is shown in curves.}
\label{fig:EoS}       
\end{figure}

When one approaches low densities, the V-QCD nuclear matter description is not reliable anymore. We therefore adopt a hybrid construction such that we demand that the holographic nuclear matter continuously transitions to ``traditional'' nuclear matter models based on effective field theories at low densities. The matching density $n\sim 1.2-2 n_s$ is left as a free parameter. Variations in this matching density, the selection of various nuclear matter models, as well as the margin in matching onto lattice QCD data of the V-QCD model at vanishing density give rise to a band in the obtained equations of state.  In figure~\ref{fig:EoS} we have described the construction \cite{Ecker:2019xrw,Jokela:2020piw}.  The blue band in figure~\ref{fig:EoS} corresponds to all those EoSs that are physically sensible while the red band follows upon taking holography into account. It is important to notice that the red band is significantly smaller (log-scale) and hence the holographic model has predictive power. This is quite remarkable and a non-trivial result as the V-QCD model is {\emph{only}} constrained at vanishing baryon densities.

\section{Applications to neutron stars}\label{sec:applications}

\begin{figure}[!ht]
\centering
\includegraphics[width=0.7\textwidth]{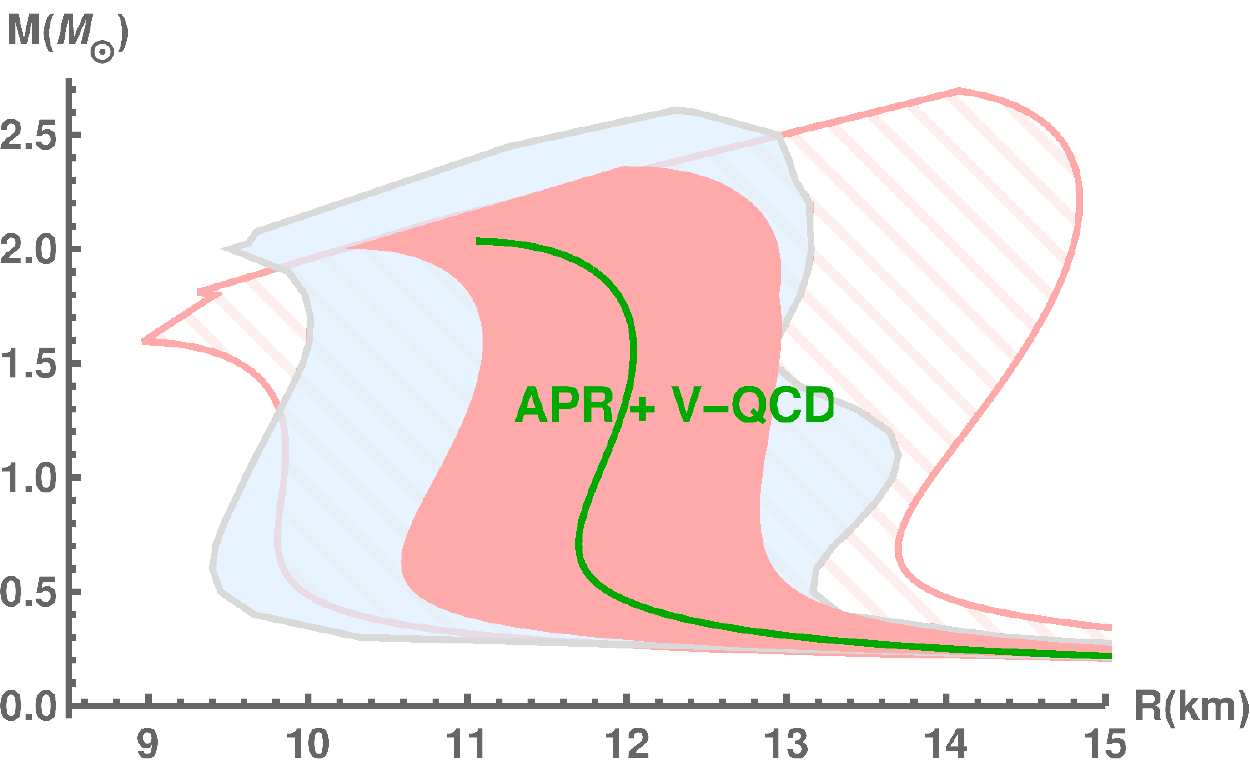}
\caption{The mass-radius curves that follow from solving the TOV equations. The color coding is the same as in figure~\ref{fig:EoS}. The V-QCD model is predictive as the red band is smaller than the blue. We have also shown the representative V-QCD(APR) curve.}
\label{fig:MRexamples}     
\end{figure}

The hybrid equations of state we have constructed suggest that indeed holography has predictive power and hence we can ask if one additionally also takes into account astrophysical constraints from neutron star this persists.  The answer is affirmative and will lead to interesting results. We therefore insert the hybrid EoSs in the Tolman-Oppenheimer-Volkov (TOV) equations are compute the mass-radius relationships for self-gravitating droplets in the standard, asymptotically Minkowski spacetime. The result following from this procedure is depicted in figure~\ref{fig:MRexamples}. Similar to figure~\ref{fig:EoS} the red band is significantly smaller than the blue, indicating that strongly interacting matter as described by holography is useful. 

Let us then implement the astrophysical constraints, in particular in the light of recent NICER observations \cite{Miller:2021qha,Riley:2021pdl}. We constrain the allowed regions by first of all demanding that the equations of state be stiff enough to accommodate two-solar mass stars as solutions, as those are detected \cite{Antoniadis:2013pzd,Cromartie:2019kug,Fonseca:2021wxt}. Second, the LIGO/Virgo collaboration also has provided constraints on the tidal deformability, especially constraining EoSs to be soft enough \cite{Abbott:2018exr}. These ($M\geq 2M_\odot,580\geq \Lambda(1.4M_\odot)\geq 70$) together with accommodating NICER observations lead to very narrow bands for the true mass-radius to reside. 
\begin{figure}[!ht]
\centering
\includegraphics[width=1.\textwidth]{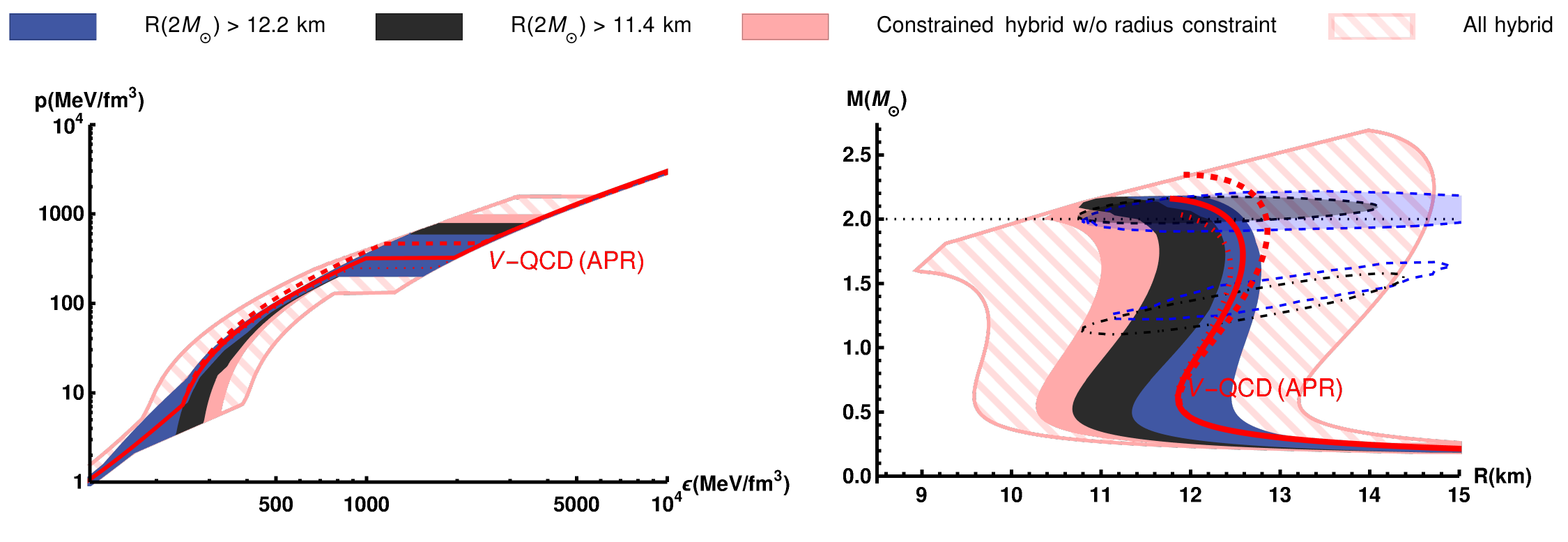}
\caption{The constrained EoS and mass-radius bands using various constraints from astrophysics. The V-QCD(APR) curves are included as representatives of the hybrid EoSs and the corresponding MR-curves.}
\label{fig:constrained}    
\end{figure}
We learn that indeed V-QCD, matched onto traditional nuclear matter models, after taking into account all current astrophysical constraints, leads to quite narrow allowed ranges. We thereby find that the model not only conforms with known physics but remains predictive. The maximum mass of the neutron stars is set by the underlying deconfinement phase transition, a phenomenon suggested in \cite{Hoyos:2016zke} to occur over a wide range of holographic models, see, {\emph{e.g.}}, \cite{Jokela:2018ers,Fadafa:2019euu,Mamani:2020pks}. This is due to the phase transition being strongly first order from the stiff nuclear matter phase to the quite soft quark matter phase. 

One of our main predictions is that the hybrid EoSs constrained with the available radius, mass, and tidal deformability measurements favor larger radii, giving tightly constrained limits $12.8$km$\geq R(1.4M_\odot) \geq 12.0$km.
Moreover, by taking the most stringent NICER bound $R(2M_\odot)>12.2$km at face value, we find that the V-QCD matched onto APR EoS at $n_{\mt{tr}}=1.6n_\mt{s}$ emerges as being an extremely nice representative of one of the most realistic equations of state to date; see figure~\ref{fig:constrained}. We have marked this curve in the figures in solid, the other dashed and dotted curves correspond to the remaining span in the high density regime, the allowed parameter range after the fit to lattice QCD.\footnote{The equations of state V-QCD(APR) can be found in the online repository CompOSE \href{https://compose.obspm.fr/}{https://compose.obspm.fr/}~\cite{Typel:2013rza}.} First numerical relativity simulations of neutron star mergers with V-QCD EoSs (matched onto SLy EoS at low density) \cite{Ecker:2019xrw} pave the road to betterment on interpreting future gravitational wave data.

\begin{figure}[!ht]
\centering
\includegraphics[width=1.\textwidth]{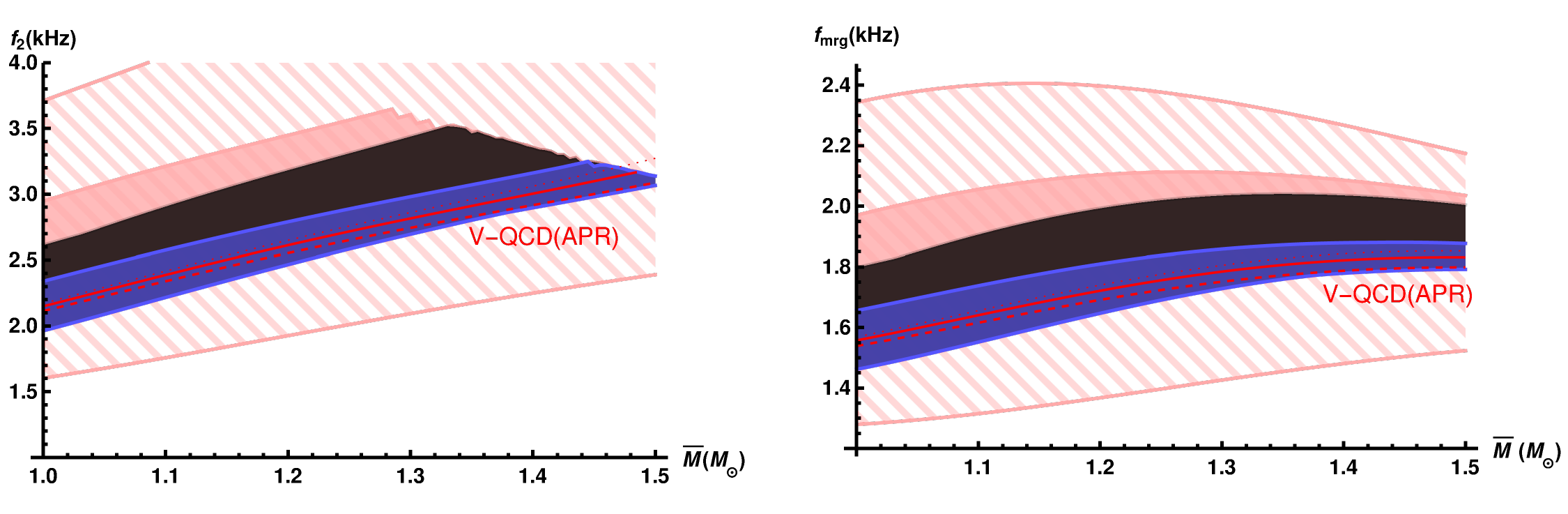}
\caption{Characteristic frequencies $f_2$ and $f_\mt{mrg}$ of the gravitational wave signal for equal mass binaries. Also shown the V-QCD(APR) as representative curves. Notice that in the left plot the prompt collapse limit cuts the bands diagonally.}
\label{fig:freqs}      
\end{figure}

In addition to static properties, we can also compute predictions for the peak frequencies of the gravitational wave signal using semi-universal relations \cite{Zappa:2017xba,Breschi:2019srl}. The characteristic frequencies of the merger and postmerger gravitational wave signal are those that correspond to the frequency at the instant of the merger and the peaks of the power spectral density of the postmerger signal, respectively. Similar to the mass-radius and EoS bands, the frequencies are tightly constrained and the clear outcome is that constraining the hybrid EoSs by radius in all of the cases is to favor lower frequencies as generic polytropic interpolations. This phenomenon is also understood by the fact that our hybrid EoSs are quite stiff. See figure~\ref{fig:freqs} for examples of the prominent peak frequency $f_2$ and the frequency at the time of the merger $f_{\mt{mrg}}$; we have chosen to present the case when there is no prompt collapse to a black hole. The frequencies are in the detectable range by LIGO/Virgo in the afterglow of the future neutron star merger events \cite{toappear}.

\section{Out of equilibrium}\label{sec:inequilibrium}

The neutron star merger is a violent event where equilibrium is not maintained and the matter heats up significantly. The simulations suggest that the temperatures can reach several tens of MeVs. This means that there is dire need also for the equation of state at finite temperature \cite{Chesler:2019osn}. This may not be enough, however, since the transport of heat and currents may become relevant also for the strongly interacting sector. Luckily, the computation of transport properties are also pretty straightforward using AdS/CFT duality, at least when the matter is in the deconfining phase.

So what are the most relevant transport properties of unpaired quark matter relevant for neutron stars? Due to the lack of systematic simulations of neutron star mergers incorporating microscopic physics beyond the equation of state this is largely unknown. See, however, \cite{Alford:2017rxf} for one example, where viscous corrections are discussed. The review \cite{Schmitt:2017efp} discusses the boundary of our understanding drawn by the effects from corrections beyond ideal hydrodynamics. The expectation is that bulk viscosity is likely the most important transport coefficient that may have a noticeable effect on the neutron star merger dynamics. The shear viscosity of the quark matter has been largely associated with the mere existence of rapidly spinning neutron stars due to the pivotal role in suppressing the r-mode instabilities. Heat and electric conductivities are relevant for the neutron star cooling and equilibration after the merger.  If the deconfinement transition is very sharp it can lead to observable signal in the gravitational wave spectrum and perhaps could be taken as a smoking gun evidence for the emergence of quark matter phase in the stellar ram \cite{Most:2018eaw,Bauswein:2018bma,Chesler:2019osn}.

While the EoS of dense and cold QCD matter has large uncertainties, even less is known about transport. Only available first-principles results for quark matter is the leading order pQCD analysis \cite{Heiselberg:1993cr} in the unpaired phase for the viscosities as well as the heat and thermal conductivities. 

The computation using the holographic approach proceeds quite straightforwardly. However, when computing the conductivities there is one stumbling block. This has to do with the fact that for a homogeneous ground state, the linear response would yield infinite results as charges accelerate indefinitely lacking any momentum dissipation. The key to finding finite results is to realize that the gradients of chemical potential and temperature can balance each other in a steady state configuration \cite{Gouteraux:2018wfe}, {\emph{i.e.}}, the transport does not occur via convection. Adapting these ideas in the context of holographic quark matter \cite{Hoyos:2020hmq,Hoyos:2021njg}, in particular the V-QCD model at hand, allows us to compare the results with those of pQCD. The results are qualitatively strikingly different and call for caution in the use of the perturbative results in neutron-star applications.

\section{Conclusions}\label{sec:conclusions}

In this talk we showed that gauge/gravity duality, combined with other approaches, is useful to study dense QCD. Using the holograpic V-QCD model with simple approximations, many details work really well. For example, we obtained a precise fit to lattice thermodynamics at vanishing baryon chemical potential $\mu\approx 0$. The model was then extrapolated to finite density and small temperatures. The resulting equation of state is still reasonable and the model has predictive power. This model is one rare example where both nuclear and quark matter phases can be treated in a unified manner.

We constructed a hybrid equation of state, where we assumed traditional nuclear matter models based of effective field theories at low densities and then we swiftly transformed to the holographic description at densities where nucleons have significant overlap. The resulting equation of state thereafter for the holographic nuclear matter phase was quite stiff while that for the quark matter phase for soft. This lead to the observation that, as a whole, all astrophysical data is consistent with our approach. One interesting conclusion is that neutron stars do not have quark matter cores, contrary to the implications in \cite{Annala:2019puf}. The reason for this is that the polytropic index for the holographic nuclear matter reaches very small values  before the deconfinement phase transition -- such small polytropic index values are classified as quark matter in \cite{Annala:2019puf}.

We also showed that the holographic framework is able to provide predictions that could be tested in future measurements. To this end we outlaid results for peak frequencies of gravitational wave spectrum. We also showed that transport properties of quark matter are quite distinct from those that one would naively obtain by extrapolating the perturbative results.

{\bf{Acknowledgements}} We would like to thank the organizers of the (virtual) Quark Confinement and the Hadron Spectrum 2021
conference in Stavanger, Norway, for the invitation to present this talk. The author has been supported in part by the Academy of Finland grant no. 1322307.

\bibliographystyle{JHEP}
\bibliography{refs.bib}

\end{document}